\documentclass[APM,showpacs,superscriptaddress,floatfix,citeautoscript,twocolumn]{revtex4-1}

\usepackage{dcolumn}
\usepackage{bm}
\usepackage{amsmath,amssymb}
\usepackage{times}
\usepackage{float}
\usepackage{graphicx, epstopdf}
\usepackage{color}
\usepackage{setspace}
\usepackage{algorithm2e}
\usepackage{notoccite}
\usepackage{siunitx}
\usepackage{natbib}

\begin{document}

\title{Hybrid nanodiamond-YIG systems for efficient quantum information processing \\
and nanoscale sensing}

\author{P. Andrich}
\affiliation{Institute for Molecular Engineering, University of Chicago, Chicago, Illinois 60637, USA}
\author{C. F. de las Casas}
\affiliation{Institute for Molecular Engineering, University of Chicago, Chicago, Illinois 60637, USA}
\author{X. Liu}
\affiliation{Institute for Molecular Engineering, University of Chicago, Chicago, Illinois 60637, USA}
\author{H. L. Bretscher}
\affiliation{Institute for Molecular Engineering, University of Chicago, Chicago, Illinois 60637, USA}
\author{J. R. Berman}
\affiliation{Institute for Molecular Engineering, University of Chicago, Chicago, Illinois 60637, USA}
\author{F. J. Heremans}
\affiliation{Institute for Molecular Engineering, University of Chicago, Chicago, Illinois 60637, USA}
\affiliation{Materials Science Division, Argonne National Laboratory, Argonne, Illinois 60439, USA}
\author{P. F. Nealey}
\affiliation{Institute for Molecular Engineering, University of Chicago, Chicago, Illinois 60637, USA}
\affiliation{Materials Science Division, Argonne National Laboratory, Argonne, Illinois 60439, USA}
\author{D. D. Awschalom}
\affiliation{Institute for Molecular Engineering, University of Chicago, Chicago, Illinois 60637, USA}
\affiliation{Materials Science Division, Argonne National Laboratory, Argonne, Illinois 60439, USA}

\begin{abstract}

The nitrogen-vacancy (NV) center in diamond has been extensively studied in recent years for its remarkable quantum coherence properties that make it an ideal candidate for room temperature quantum computing and quantum sensing schemes. However, these schemes rely on spin-spin dipolar interactions, which require the NV centers to be within a few nanometers from each other while still separately addressable, or to be in close proximity of the diamond surface, where their coherence properties significantly degrade. Here we demonstrate a method for overcoming these limitations using a hybrid yttrium iron garnet (YIG)-nanodiamond quantum system constructed with the help of directed assembly and transfer printing techniques. We show that YIG spin-waves can amplify the oscillating field of a microwave source by more than two orders of magnitude and efficiently mediate its coherent interactions with an NV center ensemble. These results demonstrate that spin-waves in ferromagnets can be used as quantum buses for enhanced, long-range qubit interactions, paving the way to ultra-efficient manipulation and coupling of solid state defects in hybrid quantum networks and sensing devices.

\end{abstract}

\maketitle 

\section{Introduction}
\vspace{-0.3cm}

Remarkable advancements have recently been made in the use of paramagnetic defects in semiconductors for quantum information and quantum sensing applications\cite{1,2,3,4,5,6}, laying the foundation for the next generation of computing machines and nanosensors. In particular, point defects in diamond and silicon carbide have attracted considerable attention because of their optical addressability and long quantum coherence persisting at and above room temperature. However, challenges remain in their adoption as qubits in functional quantum information processing (QIP) devices and quantum sensors. Many QIP applications\cite{7,8} rely on dipolar interactions between the qubits, requiring them to be within a few nanometers from each other while still allowing for their individual manipulation with optical and microwave fields. Similarly, sensing applications are based on the detection of the dipolar fields generated by target spins external to the diamond lattice, calling for the qubits to be positioned within a few nanometers from the surface, where their coherence properties are strongly diminished\cite{9,10}.

Hybrid ferromagnet-solid state qubit quantum systems have recently been proposed as an architecture with great potential to tackle some of these limitations by introducing quantum buses between distant qubits\cite{11}, or by enhancing the sensitivity capabilities of quantum sensors\cite{12}. In this work we use spin-waves (SWs) excited in a yttrium iron garnet (YIG) thin film to mediate long-range coherent interactions between a microwave source and nitrogen-vacancy (NV) centers in nanodiamonds (NDs), demonstrating a fundamental step towards obtaining interconnected quantum networks and quantum sensing devices based on paramagnetic defects in the solid state. We show, in particular, that surface Damon-Eshbach spin-wave (DESW) modes excited in the YIG film result in the amplification of a microwave field created by a microstrip line (MSL) antenna by more than two orders of magnitude. We use this amplification to obtain SW mediated resonant driving of NV centers located in an area hundreds of microns wide and to demonstrate highly efficient coherent control using a power that is over three orders of magnitude lower than what is necessary using the microwave field generated directly by the antenna. To accurately determine the origin and extent of the observed phenomena we rely on a novel transfer printing technique that enables the precise control of the NDs position. The strong microwave field amplification and the coherent nature of the SW-NV center interactions demonstrate the possibility of relaxing the distance requirements imposed by direct dipolar coupling and enhancing the NV centers sensitivity by using ferromagnetic thin films as quantum buses for inter-qubits and qubit-target spin connections.

\begin{figure*}[t]
\centering
\includegraphics{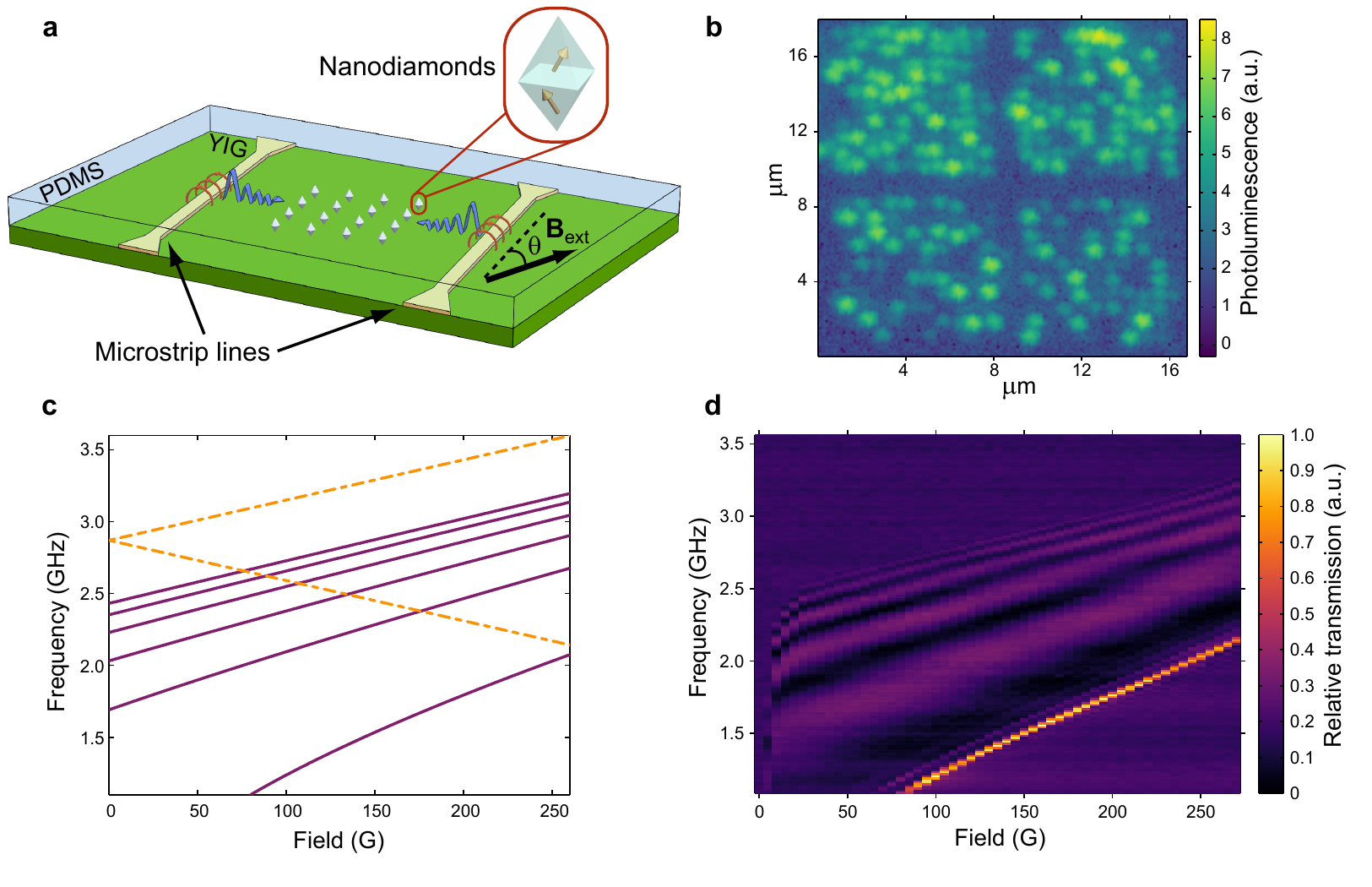} 
\caption{
Setup and spin-wave dispersion relation of the system. \textbf{a}, Sample schematics. The sample is a \SI{3.08}{\micro\meter} thick single-crystal YIG film with a \SI{\sim 300}{\micro\meter} thick PDMS strip laid on top. The PDMS layer contains an array of NDs that are in contact with the YIG substrate. Two \SI{5}{\micro\meter} wide microstrip lines (MSL) are patterned \SI{100}{\micro\meter} apart on the YIG to apply microwave fields. The microwave magnetic fields (circles) and the direction of the propagating spin-waves are indicated. An external magnetic field (Bext) is applied at an angle $\theta$ with respect to the MSLs. The arrows in the ND represent the NV center spins, with each ND containing hundreds of NV centers. \textbf{b}, Spatial photoluminescence scan of ND arrays. \textbf{c}, Simulated spin wave spectrum for our sample in the case of magnetic field parallel to the MSLs. The dashed lines enclose the range of frequencies where the NV centers’ ground state spin resonances lay. \textbf{d}, Microwave transmission spectrum between the MSLs as a function of the externally applied magnetic field for $\theta = 0$. The data for $0$ gauss was subtracted from the data at higher fields to eliminate features that are not field dependent. 
 }
\end{figure*}

\vspace{-0.5cm}
\section{Experimental apparatus}
\vspace{-0.2cm}

To investigate the SW properties and their interactions with NV centers we use the setup shown in Fig.\nobreakspace1a. A pair of Ti/Au MSLs \SI{5}{\micro\meter} wide and \SI{200}{\nano\meter} thick, separated by \SI{100}{\micro\meter} is lithographically patterned on the surface of a \SI{3.08}{\micro\meter} thick YIG layer epitaxially grown on a gadolinium gallium garnet substrate (GGG). The MSLs are tilted with respect to the sample edges to avoid SW reflections. YIG is chosen as a substrate because of its small damping parameter for spin wave propagation in the GHz frequency range\cite{13}, which makes it ideal for studying long-range interactions. Positioned in contact with the YIG layer is an array of commercial NDs (Adamas Technology, $\sim500$ NV centers per particle) embedded on the surface of a \SI{\sim 300}{\micro\meter} thick strip of polydimethylsiloxane (PDMS), which was fabricated through chemical pattern directed assembly\cite{14} of NDs on a silicon substrate followed by transfer printing\cite{15} with PDMS as described in Supplementary Section 2. This portable and reusable system allows us to control the position of the NDs with respect to the MSLs and to easily locate and address single nanoparticles, which is critical for interpreting our measurements. Additionally, the flexibility of the PDMS guarantees the presence of close contact between the NDs and the YIG substrate. In Fig.\nobreakspace1b we show a photoluminescence (PL) spatial scan of a typical section of the ND array as collected through a custom-built confocal microscopy apparatus (Supplementary Section 2).

\begin{figure*}[t]
\centering
\includegraphics{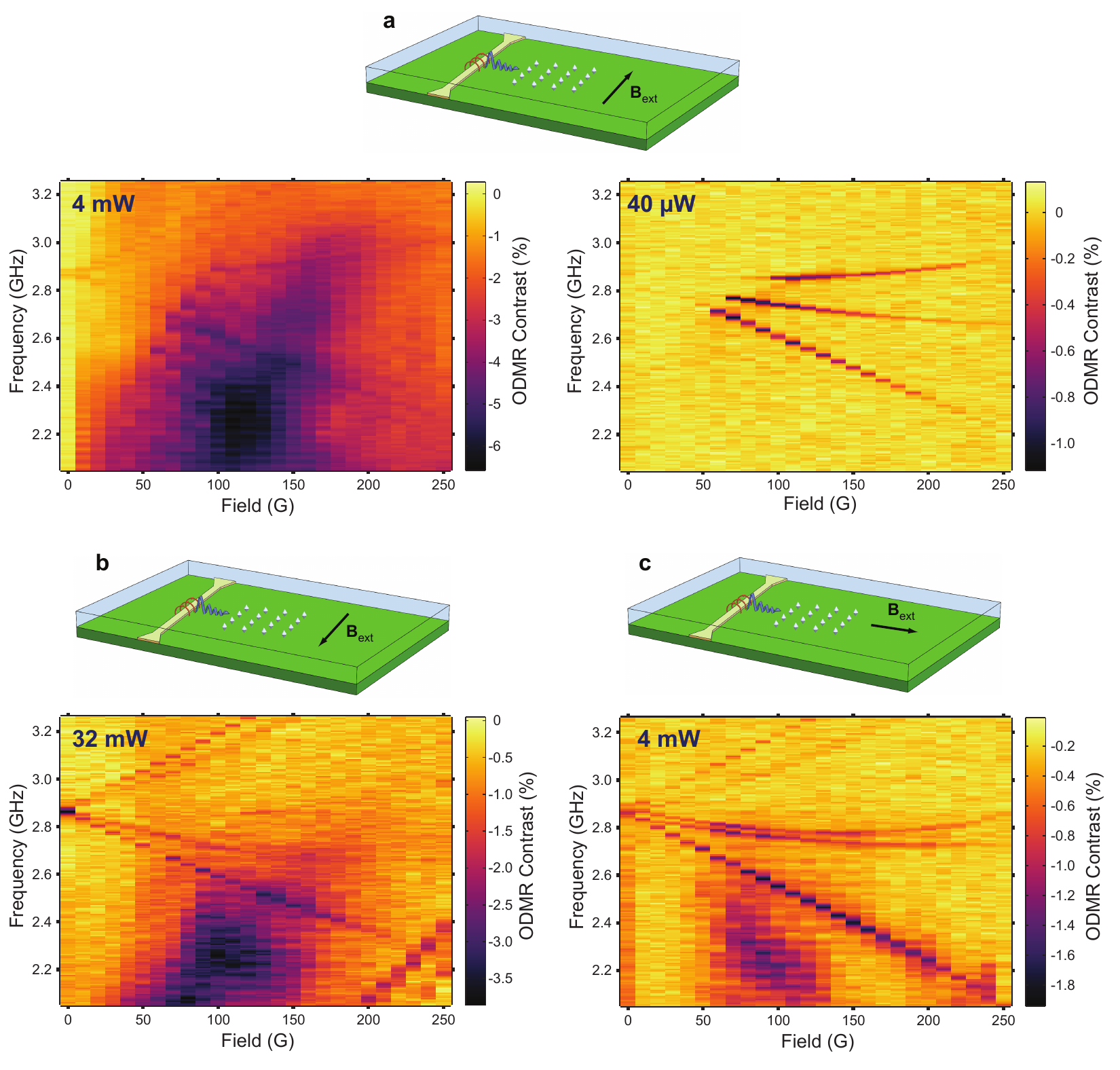} 
\caption{
ODMR spectra of NV centers in NP-P for different orientations of the external magnetic field and microwave power. \textbf{a}, Spectra collected using \SI{4}{\milli\watt} and \SI{40}{\micro\watt} in the case of $\theta = 0$. \textbf{b,c}, Spectra collected for $\theta = \pi$ using \SI{32}{\milli\watt} and for $\theta = \pi$/2 using \SI{4}{\milli\watt} respectively.
 }
\end{figure*}

\begin{figure*}[!t]
\centering
\includegraphics{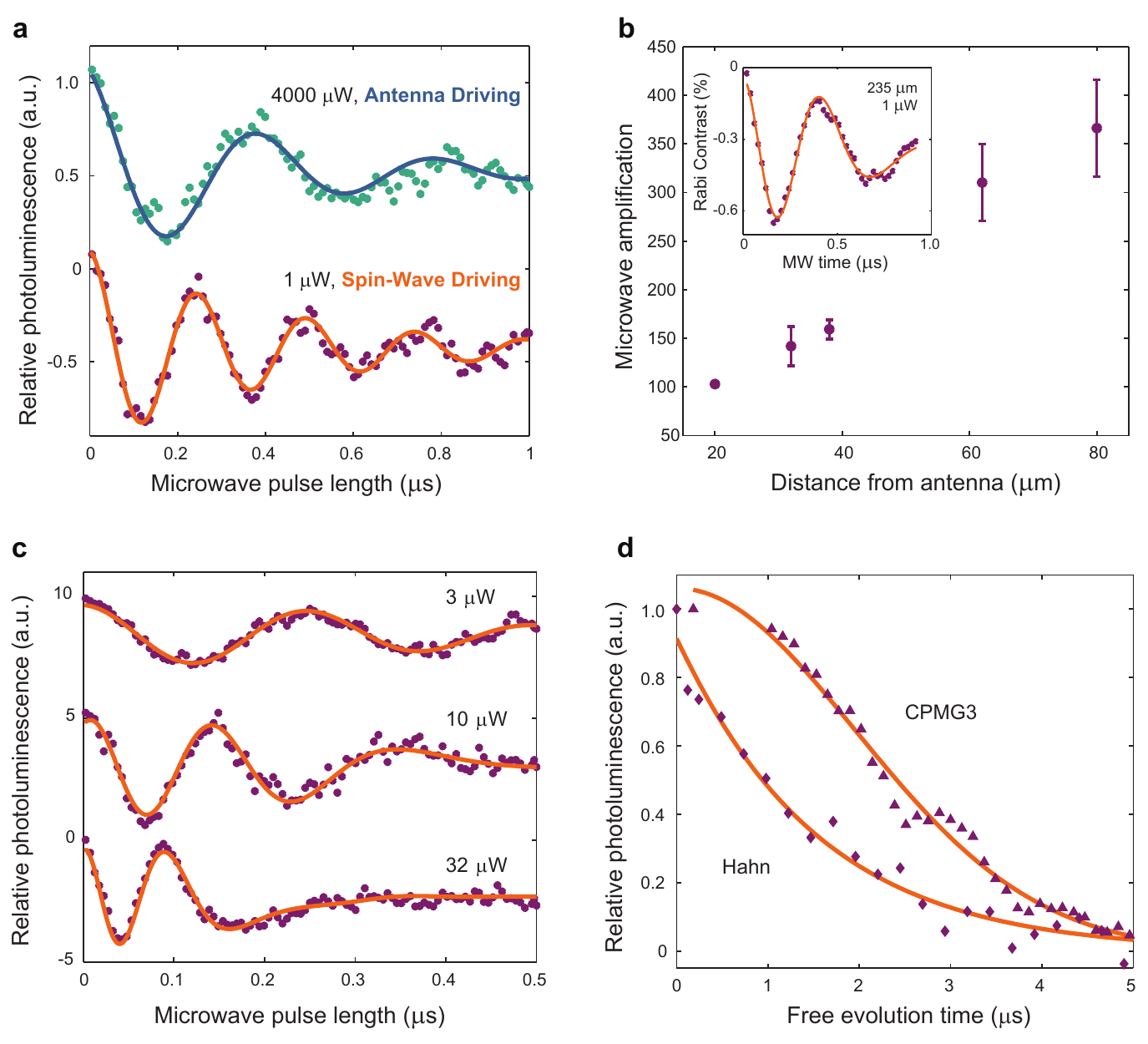} 
\caption{
Time resolved measurements and spatial dependence of the microwave field amplification. \textbf{a}, Rabi oscillations measured on NP-Q (\SI{20}{\micro\meter} away from the MSL) at the same microwave driving frequency (\SI{2.862}{\giga\hertz}) but different external magnetic field and microwave power. The top curve is measured in the antenna driving regime at $15$ gauss and using  (\SI{4}{\milli\watt}) of microwave power while the bottom curve is collected in the pure SW driving regime at $145$ gauss and using \SI{1}{\micro\watt} of microwave power. \textbf{b}, Microwave field amplification obtained when using SW driving (high fields) over antenna driving (low fields) as a function of the ND’s distance from the antenna. The error bars reflect imprecisions in the determination of the Rabi frequencies. In the inset we show the Rabi signal obtained at high fields when the particle is \SI{\sim 235}{\milli\meter} away from the antenna using \SI{1}{\micro\watt} of microwave power. \textbf{c}, Rabi oscillations measured on NP-Q at a fixed external magnetic field ($120$ gauss) and different microwave powers. \textbf{d}, Hahn-echo and CPMG3 pulsed measurements that show robust multi-pulse control of the NV centers. Both sets of data are renormalized and fit to exp[-(t/T2)$^{\alpha}]$, where $\alpha = 1$ and $2$ for the Hahn and CPMG3 case respectively. From these fits we obtain T$_{2,Hahn}$ = \SI{1.54}{\micro\s} and T$_{2,CPMG3}$ = \SI{2.78}{\micro\s}.
 }
\end{figure*}

\vspace{-0.5cm}
\section{Spin-wave spectrum}
\vspace{-0.2cm}

As we are interested in the effects of the resonant interactions between SWs and NV centers we first calculate and experimentally measure the SW spectrum of our system to ascertain where it overlaps with the NV center's spin resonances. While the direction of SW propagation is always orthogonal to the SML, SW modes with different dispersion relations and magnetization profiles across the ferromagnetic layer can be excited depending on the relative orientation of the externally applied magnetic field B and the direction of propagation of the SWs\cite{16}. Here we primarily focus on DESWs (unless otherwise stated), which are excited when the external magnetic field is in the plane of the YIG film and parallel to the MSL ($\theta = 0$). We select these modes as their energies lie closest to the NV center spin ground state transitions at the magnetic fields used in this work (B $= 0$ to $250$ gauss) and their surface nature could provide the strongest interaction with external spins. We calculate the theoretical spectrum of the DESW following the approach detailed in Supplementary Section 3 and report the result of this calculation in Fig.\nobreakspace1c. In the same figure (dashed lines) we identify the frequency range for the NV center spin ground state resonances, which is enclosed by the resonance spectrum for a defect aligned with the external magnetic field. Even though the nanoparticles have random crystal orientations, their resonances inherently fall within this range\cite{17}.  

We also experimentally measure the SW modes dispersion using microwave transmission measurements between the two MSLs\cite{18} (Supplementary Section 4). The zero-field measurement is used as a reference for the ones at higher fields to eliminate the features in the spectrum that are not magnetic field dependent. The data in Fig.\nobreakspace1d shows good agreement with the calculated spectrum. In order to interpret the results described in what follows it is important to note that DESW excitations with higher frequencies (at a fixed external field) are associated with larger wave vectors \textit{k} (Supplementary Section 3), and display a more pronounced surface confinement as both the magnetization oscillations within the ferromagnetic layer and the field generated outside decay exponentially with characteristic length \textit{1/k} in the direction orthogonal to the YIG surface\cite{16}. 

\vspace{-0.5cm}
\section{Optically detected resonant interaction}
\vspace{-0.2cm}

To study the extent of the DESW-NV centers interaction, we perform optically detected magnetic resonance (ODMR) measurements using one of the MSLs to drive microwave fields. Fig.\nobreakspace2a shows the magnetic field configuration used for these measurements and the ODMR spectra obtained on a nanoparticle labeled NP-P, located  \SI{\sim 40}{\micro\meter} away from the MSL. On the left we report the results obtained using \SI{\sim 4}{\milli\watt}  of microwave power, which is the lowest power that still resolves the NV center`s resonances at low fields. When we increase the external magnetic field we observe a broad feature of increasing frequency that intersects the NV center resonances following the SW resonances' expected behavior. Recent studies have shown the presence of less prominent off-resonant features in the NV centers` ODMR spectrum\cite{19,20} ascribed to the shortening of the NV center longitudinal (T$_{1}$) spin coherence time caused by the broad spectrum magnetic field noise introduced by the excited spin waves. Because of the extensive quenching effect on the spin coherence of the NV centers, it is not possible to isolate the effect of resonant interactions between the SWs and the NV centers at this microwave power level.

On the right side of Fig.\nobreakspace2a, we present the ODMR spectrum obtained using \SI{40}{\micro\watt} of microwave power. Under this condition, the off-resonant effects are not visible but some of the NV center ODMR features are still prominent in the field-frequency subspace where they cross the spin wave resonances. We stress that at magnetic fields below $50$ gauss, no ODMR is visible due to the microwave magnetic field generated by the antenna being insufficiently strong to directly drive the NV centers. Only at higher fields does the microwave antenna create spin waves that strongly interact with the NV centers at its resonant frequency, creating optical contrast. This result clearly indicates the presence of a purely SW driven excitation. We note that only the lower branches of the NV centers’ ground state spin transitions are visible. We attribute this phenomenon in part to the fact that the upper branches are not intersected by the SW resonances, as can be inferred from Fig.\nobreakspace1b. Additionally, the transmission of the MSLs also decreases at higher frequencies resulting in a lower excitation efficiency. While we cannot resolve the effect of separate SW excitations, this is due to the coarse magnetic field steps used here ($10$ gauss). When finer steps are taken, the discrete nature of the SW spectrum becomes clearly visible (Supplementary Section 6). Moreover, we note in Fig.\nobreakspace2a that the interaction appears to be stronger for the SWs associated with larger wave vectors, as can be deduced from the decrease in ODMR contrast at higher fields, where the NV centers’ resonances cross lower \textit{k} modes. This is consistent with a stronger surface confinement of the magnetization oscillations for higher values of \textit{k}.

We investigate the effect of the magnetic field orientation on the SW-NV resonant interaction for the cases $\theta$ = $\pi$ and $\theta$ = $\pi$/2. In these conditions the excited spin waves have different dispersion relations and magnetization profiles, which allows us to analyze the dependence of the SW-NV interactions on these properties. In Fig.\nobreakspace2b we show the ODMR data collected for the $\theta$ = $\pi$ case using \SI{32}{\milli\watt} of microwave power, which is the minimum power needed to resolve the NV centers resonances at all fields. The strong reduction in the PL quenching can be explained in light of the non-reciprocal nature of the DESW modes\cite{21}, which implies that a $\theta$ = $\pi$ rotation of the magnetic field results in a decrease of the SWs’ excitation efficiency\cite{22,23} and in a drastic change in the SWs amplitude profile, which is confined to the opposite surface of the ferromagnetic layer\cite{24}. In the case of $\theta$ = $\pi/2$, pure backward volume magnetostatic spin waves (BVMSW) are excited\cite{16,24}. These modes have lower resonant frequencies (Supplementary Section 5) than the DESW and are characterized by a sinusoidal magnetization oscillation profile across the thickness of the ferromagnetic layer.  In Fig.\nobreakspace2c we present the ODMR spectrum collected using \SI{4}{\milli\watt} of microwave power. The extent to which the PL is affected is remarkably smaller than for the $\theta$ = 0 case. While the excitation efficiency for DESW and BVMSW is different, this alone cannot explain the two orders of magnitude increase in power required to observe ODMR contrast in the latter case\cite{25}. The difference in the frequencies of the two sets of modes also does not justify the absence of a region of strong PL quenching in the magnetic field range we studied, particularly considering the broadband nature of the off-resonant effects. Together, the measurements presented in Fig. 2 demonstrate the ferromagnetic nature of the enhanced microwave-NV center interactions and that the surface confinement of DESW greatly contributes to this enhancement.

\vspace{-0.2cm}
\section{Hybrid coherent driving of NV centers}
\vspace{-0.2cm}

We establish the viability of hybrid YIG-ND systems for quantum information and sensing applications by demonstrating coherent Rabi driving of NV centers using DESWs. By measuring Rabi oscillations on multiple NDs at magnetic fields below and above the threshold for SW-NV interaction, we determine that the SWs can be the source of coherent control and quantify the enhancement that the SW driving provides over using the microwave field generated by the antenna. In contrast to previous demonstrations of off-resonant interactions between YIG and NDs\cite{19,20}, these results show that SWs can serve as long range buses for coherent microwave signals without quenching the quantum coherence of the NV centers. In order to isolate the effect of the SW mediated driving from other parameters that influence the SW-NV interaction, such as the frequency dependence of microwave power transmission of the MSL and the NV center orientation, we focus on a nanoparticle (NP-Q) that contains NV centers aligned nearly perpendicularly to the external magnetic field. For these NV centers, the lower branch of the ground state spin transition can assume the same energy at low and high fields (Supplementary Section 7), which allows us to compare measurements collected from the same subset of the NV center ensemble using the same microwave frequency. In Fig.\nobreakspace3a we report the data collected using a driving field resonant with the analyzed NV transition both at $15$ and $145$ gauss, using \SI{4}{\milli\watt} and \SI{1}{\micro\watt} of power respectively. These measurements clearly show that it is possible to coherently drive the NV centers exclusively using interactions with the propagating SW. Moreover, this data shows a SW mediated amplification of the driving microwave magnetic field by a factor of $\sim$100 (Supplementary Section 8). This amplification factor is a function of the nanoparticle’s distance from the antenna because, while the antenna’s field sharply decays as the inverse of the distance from it, the magnetic field generated by the propagating spin waves is only limited by the YIG’s large spin wave propagation length. This effect is illustrated in Fig.\nobreakspace3b where we show the behavior of the microwave field amplification for NP-Q as a function of its distance from the antenna. When we translate the ND from $20$ to $80$ \si{\micro\meter} away from the MSL, the microwave magnetic field amplification increases roughly linearly to $> 350$, suggesting that the SW decay length in the YIG substrate is significantly larger than \SI{80}{\micro\meter}. To illustrate this point and to showcase the long-range nature of the SW-NV centers interaction we show the Rabi measurement (inset of Fig.\nobreakspace3b) collected \SI{\sim 235}{\micro\meter} away from the antenna using \SI{1}{\micro\watt}  of microwave power. We note that at distances \SI{> 80}{\micro\meter}, it was not possible to observe direct driving induced by the antenna's electromagnetic field because of the limited available microwave power making it impossible to determine the amplification factor. 

The effect of the microwave power on the SW mediated coherent driving is portrayed in Fig.\nobreakspace3c, where we show a series of measurements collected on NP-Q. The Rabi frequency increases linearly with the square root of the input power in the range used in this work, even in the pure SW driving regime (Supplementary Section 9), suggesting that we can neglect the impact of nonlinear effects. This allows us to associate the effect described in Fig.\nobreakspace3b only to the difference in the spatial profiles of the antenna and SW fields, independently of the microwave power used for each measurement. In the data in Fig.\nobreakspace3c, we also notice a reduction in the temporal extent of the Rabi oscillations with increasing microwave powers. This phenomenon is likely the result of an increase in the non-resonant magnetic noise (see effect of higher microwave power on the ODMR data) that adversely affects the coherences of the NV centers, and can be counteracted by using higher quality NDs with controlled geometry and NV center density, which possess longer coherence times\cite{26}. 

We further demonstrate the robustness of the spin wave mediated coherent control using advanced multi-pulse dynamical decoupling protocols that are the basis for sensing and quantum computing applications. In Fig.\nobreakspace3d we show the result obtained in the pure SW driving regime for an additional nanoparticle (NP-R) \SI{\sim 70}{\micro\meter} away from the MSL using \SI{\sim 5}{\micro\watt} of microwave power. The ability to extend the coherence time using multi-pulse sequences demonstrates full control of the NV centers through the pure SW driving. We note that, at the microwave power levels required to perform these measurements, the spin coherence times are not significantly altered by the presence of the ferromagnet, as it is clear from measuring T2 on a ND first while in contact with YIG and then with a non-ferromagnetic GGG substrate (Supplementary Section 10). 

\vspace{-0.3cm}
\section{Enhanced Sensing with Microwave Magnetic Field Amplification }
\vspace{-0.2cm}

\begin{figure}[t]
\centering
\includegraphics{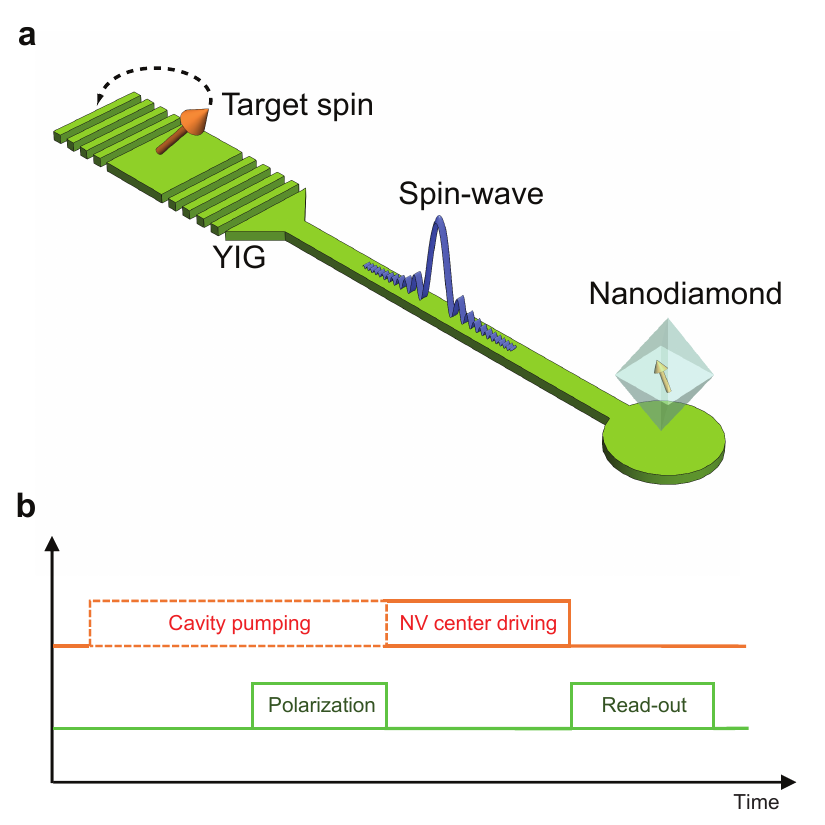} 
\caption{
Conceptual representation of a hybrid sensing scheme based on remote spin-wave driving of NV centers. \textbf{a}, Schematic of the proposed design for the sensing device. \textbf{b}, Pulse sequence for the detection of electronic spins through indirect coherent driving of the NV centers. 
 }
\end{figure}

The results we present suggest that hybrid YIG-ND systems can be useful for enhancing the NV center sensing capabilities by relaxing the limits imposed by the NV-target dipolar interactions. To illustrate this concept, we propose a device (Fig.\nobreakspace4a) in which target spins (e.g. free radicals in a biological molecule) are placed on top of a magnonic cavity that is connected by a SW waveguide to a distant ND. Following the pulse sequence shown in Fig.\nobreakspace4b, a microwave field resonant with the target spins’ Zeeman transition but not with the SW cavity mode, which is resonant with a NV center’s transition, induces Rabi oscillations of the target spins. When the periodicity of these oscillations matches the frequency of the cavity mode, the latter is progressively pumped by the microwave field generated by the precessing target spins, until the rate of SWs leaking from the cavity matches the rate of pumping. The NV centers are then initialized using a laser pulse, before interacting with the spin waves leaking out of the cavity for a variable time $\tau$. Finally, the state of the NV centers is read out optically with a second laser pulse. This measurement scheme is equivalent to performing SW mediated Rabi driving on the NV centers and contains information on the species of the target spins through the resonant nature of their driving, and on their concentration through the Rabi oscillations frequency for a fixed pumping time. We note that similar systems can be developed to obtain long distance interactions between NV centers\cite{11}. 

\vspace{-0.3cm}
\section{Conclusion}
\vspace{-0.2cm}

We demonstrate the use of a hybrid YIG-ND system to obtain long-range, purely SW mediated coherent control of diamond NV centers. We show that propagating surface SWs in a YIG thin film can locally amplify an antenna driven microwave field by more than two orders of magnitude, and this amplification persists up to hundreds of micrometers away from the microwave source and it is only limited by the SW propagation length. Additionally, we demonstrate the viability of using strong SW-NV center coherent interaction to implement advanced dynamical decoupling schemes. Further enhancement of this interaction can be achieved by using an antenna designed\cite{27} to create higher wave-vector SW modes with even more surface confinement.  The use of engineered NDs with controllable NV center density, position, and orientation, as well as much improved coherence times\cite{26}, could greatly improve the use of the YIG-ND platform for quantum information and sensing applications while remaining compatible with the flexible PDMS membrane used here. The recent advent of spin-wave waveguides\cite{28,29} and cavities\cite{30} composed of microfabricated YIG films also suggests that separate elements within a nanoparticle array could be addressed individually with a local YIG based microwave source for applications where a global microwave antenna is not ideal. Finally, the demonstration of low power control of solid state paramagnetic defects illustrates the potential of hybrid ferromagnet-qubit platforms as promising pathways for energy efficient classical and quantum spintronics devices. 

The authors thank B.B. Zhou and M. Fukami for useful discussions. This work was supported by the Army Research Office through the MURI program W911NF-14-1-0016 and U.S. Air Force Office of Scientific Research FA8650-090-D-5037. P.F.N., F.J.H. and D.D.A. were supported by the US Department of Energy, Office of Science, Basic Energy Sciences, Materials Sciences and Engineering Division.

\end{document}